\documentclass[12pt]{article}

\usepackage[fleqn]{amsmath}
\usepackage{amsfonts}
\usepackage{amssymb}
\usepackage{epsfig}
\usepackage{geometry}
\usepackage{ctable}
\usepackage{float}

\newcommand{\bE}{\mathbf{E}}
\newcommand{\bK}{\mathbf{K}}

\newcommand{\drm}{\mathrm{d}}

\newcommand{\refeq}[1]{(\ref{#1})}

\newcommand{\vect}[1] {\boldsymbol{{ #1}} }

\newcommand{\Nset}{\mathbb{N}}

\newcommand{\Rset}{\mathbb{R}}

\newcommand{\kV}{{\vect{k}}}		% 3-current density
\newcommand{\qV}{{\vect{q}}}            % 3-position of particle
\newcommand{\sV}{{\vect{s}}}            % position 3-vector

\newcommand{\BV}{\pmb{{\cal B}}}

\newcommand{\DV}{\pmb{{\cal D}}}
\newcommand{\EV}{\pmb{{\cal E}}}

\newcommand{\HV}{\pmb{{\cal H}}}

\newcommand{\alphaS}{\alpha_{\mbox{\tiny{S}}}}

\newcommand{\mEL}{{m}_{\text{\textrm{el}}}}       % electron mass
\renewcommand{\leq}{\leqslant}
\renewcommand{\geq}{\geqslant}

\reversemarginpar

\numberwithin{equation}{section}

\begin{document}

%%%%%%%%%%%%%%%%%%%%%%%%%%%%%%%%%%%%%%%%%%%%%%%%%%%%%%%%%%%%%%%%%%%%%
%%%%%%%%%%%%%%%%%%%%%%%%%%%%%%%%%%%%%%%%%%%%%%%%%%%%%%%%%%%%%%%%%%%%% 
%       
\title{\uppercase{Do particles and anti-particles  \\ really annihilate each other?}}
\author{\textbf{Michael K.-H. Kiessling}\\
          Department of Mathematics, Rutgers University\\
                110 Frelinghuysen Rd., Piscataway, NJ 08854, USA}
\date{\textrm{\small Version of October 05, 2018}} % \\ Submitted to Phys. Lett. B.}}

\vspace{-1truecm}
\maketitle

\thispagestyle{empty}

\vspace{-1truecm}
\begin{abstract}
{\noindent 
 Supported by results obtained with semi-classical quantization techniques, and with a 
quantum mechanical ``square root Klein--Gordon'' operator, it is argued that positronium (Ps) may exhibit a proper 
quantum-mechanical ground state whose energy level lies $\approx 2\mEL c^2$ below its ``hydrogenic pseudo-ground state'' energy,
where $\mEL$ is the empirical rest mass of the electron.
 While the familiar hydrogenic pseudo-ground state of Ps is caused by the electric attraction of electron and anti-electron, modified by small
magnetic spin-spin and radiative QED corrections, the proper ground state of Ps is caused by the magnetic attraction
of electron and anti-electron, which dominates the electric one at short distances.
 This finding suggests that the familiar ``annihilation'' of electron and anti-electron is, in reality, simply yet another 
transition between two atomic energy levels, with the energy difference radiated off in form of photons --- except that the 
energy difference is huge: about 1 MeV instead of the few eV in a hydrogenic transition. 
 In this proper ground state configuration the two particles would be so close that they would electromagnetically 
neutralize each other for most practical purposes, resulting in the appearance that they have annihilated each other.
 Once in such a tightly bound state such pairs would hardly interact with normal everyday matter and not be
noticeable --- except through their gravitational effects in bulk.
 If the existence of such a low-energy ground state is confirmed, it would imply that
a significant part of the mysterious ``dark matter'' in the universe may consist of such matter-antimatter bound states.}
\end{abstract}

\vfill

% An earlier preprint is available at RGate; DOI:  10.13140/RG.2.2.22279.70566}

\medskip
\hrule
\smallskip
\copyright(2018) \small{The author. Reproduction of this preprint, in its entirety, and 

\hspace{1.6truecm} for non-commercial purposes only, is permitted.}

\newpage

%%%%%%%%%%%%%%%%%%%%%%%%%%%%%%%%%%%%%%%%%%%%%%%%%%%%%%%%%%%%%%%%%%%%%%%%%%%

%\baselineskip=24pt

%%%%%%%%%%%%%%%%%%%%%%%%%%%%%%%%%%%%%%%%%%%%%%%%%%%%%%%%%%%%%%
%%%%%%%%%%%%%%%%%%%%%%%%%%%%%%%%%%%%%%%%%%%%%%%%%%%%%%%%%%%%%%
%%%%%%%%%%%%%%%%%%%%%%%%%%%%%%%%%%%%%%%%%%%%%%%%%%%%%%%%%%%%%%
                \section{Introduction}\vspace{-10pt}
%%%%%%%%%%%%%%%%%%%%%%%%%%%%%%%%%%%%%%%%%%%%%%%%%%%%%%%%%%%%%%
%%%%%%%%%%%%%%%%%%%%%%%%%%%%%%%%%%%%%%%%%%%%%%%%%%%%%%%%%%%%%%
%%%%%%%%%%%%%%%%%%%%%%%%%%%%%%%%%%%%%%%%%%%%%%%%%%%%%%%%%%%%%%
\noindent
 In this paper we will use semi-classical quantization techniques, and quantum mechanical operator estimates,
to supply theoretical evidence for the viability of the idea that what appears to be the annihilation of a matter-antimatter 
pair of particles may in reality simply be a transition of the two particles into an extremely tightly bound quantum state with 
energy level $\approx 0$, i.e. essentially $2mc^2$ below the positive energy continuum of this system.
 The electromagnetic signature of such a transition would be largely indistinguishable from that of an actual annihilation process.
 Although the two particles would not literally annihilate each other, in such an almost zero-energy ground state their charges and 
magnetic moments would effectively neutralize each other for most practical purposes --- hence the appearance of annihilation.
 Similarly, what appears as the creation of a matter-antimatter particle pair would simply be the ionization of this two-particle atom
from its ground state, by injecting an energy of $\approx 2mc^2$ into the system, with essentially zero total momentum, viewed from
their center-of-mass (CM) frame.
 As a consequence, such a tightly bound state of particle and anti-particle would barely interact with normal matter particles.\vspace{-0.5pt}
 
 If this scenario is confirmed by properly relativistic quantum-mechanical two-body computations --- and ultimately: experiments ---, 
it would not entail the conclusion that the usual quantum field-theoretical formalism is entirely false (though its well-known 
divergence problems make it plain that this formalism is also not entirely satisfactory).
 Yet it would add another reason to believe that QFT is only an ``effective field theory'' which needs to be modified when pushed
too far. 
 In particular, ``the vacuum'' for a particle \&\ anti-particle sector, normally the stand-in for its ``ground state,'' would have to be 
replaced by a nontrivial proper ground state (a superselection rule).
The ``annihilation and creation'' operator formalism would have to be {interpreted} as describing,
not the annihilation and creation of the particles themselves, but rather their transitions between quantum states.\vspace{-0.5pt}
 
  One potentially important spinoff of our theoretical investigations is the speculation that the universe might be filled with such tightly 
bound matter plus anti-matter pairs in their proper atomic ground states. 
 While individually practically invisible, altogether they may well contribute to universal gravitation
(if the ground state energy is not exactly zero).
 Thus, a significant part of the mysterious ``dark matter'' in the uinverse may consist of such matter-antimatter bound states.
 Since these tightly bound particle \&\ anti-particle pairs would have spin 0, they are effectively bosons and may therefore provide
a vindication of the bosonic dark matter models proposed already in \cite{RuffiniETal} and more recently by others \cite{HOTW}.
 This issue will be addressed in more detail in a subsequent publication.\vspace{-0.5pt}

 The rest of the paper is structured as follows:
 In section \ref{sec:ePLUSeMINUS} we demonstrate the viability of our suggestion with a semi-classical and with a quantum-mechanical 
treatment of the quintessential matter-antimatter system: Positronium (Ps).
 In section 3 we summarize and re-inforce our findings, re-emphasizing the potential implications for our 
understanding of the nature of dark matter, and the apparent matter / anti-matter asymmetry in the universe; 
see also \cite{CERNa,CERNb} for a CERN perspective.
 In the Appendix we comment on other particle \&\ anti-particle annihilation scenarios.\vspace{-1.5pt}

\newpage

%%%%%%%%%%%%%%%%%%%%%%%%%%%%%%%%%%%%%%%%%%%%%%%%%%%%%%%%%%%%%%
%%%%%%%%%%%%%%%%%%%%%%%%%%%%%%%%%%%%%%%%%%%%%%%%%%%%%%%%%%%%%%
%%%%%%%%%%%%%%%%%%%%%%%%%%%%%%%%%%%%%%%%%%%%%%%%%%%%%%%%%%%%%%
                \section{Positronium}\label{sec:ePLUSeMINUS}
%%%%%%%%%%%%%%%%%%%%%%%%%%%%%%%%%%%%%%%%%%%%%%%%%%%%%%%%%%%%%%
%%%%%%%%%%%%%%%%%%%%%%%%%%%%%%%%%%%%%%%%%%%%%%%%%%%%%%%%%%%%%%
%%%%%%%%%%%%%%%%%%%%%%%%%%%%%%%%%%%%%%%%%%%%%%%%%%%%%%%%%%%%%%
 Positronium is a hydrogen-like two-body system composed of an electron and an anti-electron mutually bound to each other by their 
electromagnetic attraction. 
 However, different from hydrogen, Ps in its hydrogenic ground state is not stable,
 having a life span of $0.125\times10^{-3}\,\mu$s (p-Ps),
respectively of $0.14\, \mu$s (o-Ps);
here, p-Ps and o-Ps mean para-Ps and ortho-Ps, respectively, referring to whether the electron and anti-electron spins add to 0
(p-Ps) or to 1 (o-Ps). 
 The shorter-lived p-Ps can decay through emission of two photons, while o-Ps requires at least three photons to decay --- hence
its much longer life span.
 There is also a small energy difference between the hydrogenic p-Ps and o-Ps ground states, known as hyperfine (HF) 
splitting, caused mainly by their spin-spin interaction; see \cite{KarplusKlein}, \cite{positronium}, \cite{PsS}. 
 In this sense the hydrogenic p-Ps ground state \emph{is} the hydrogenic ground state of Positronium.\footnote{Since 
   the hydrogenic ground state of Ps is not stable, it is not a quantum-mechanical ground state in the proper sense.
   However, in the usual annihilation narrative, there is no lower-energy ``state of Ps.'' 
   We prefer to call the hydrogenic ground state of Ps \emph{a pseudo-ground state}.\vspace{-10pt}}
 
%%%%%%%%%%%%%%%%%%%%%%%%%%%%%%%%%%%%%%%%%%%%%%%%%%%%%%%%%%%%%%
%%%%%%%%%%%%%%%%%%%%%%%%%%%%%%%%%%%%%%%%%%%%%%%%%%%%%%%%%%%%%%
                \subsection{The hydrogenic p-Ps annihilation physics}\label{sec:pPs}
%%%%%%%%%%%%%%%%%%%%%%%%%%%%%%%%%%%%%%%%%%%%%%%%%%%%%%%%%%%%%%
%%%%%%%%%%%%%%%%%%%%%%%%%%%%%%%%%%%%%%%%%%%%%%%%%%%%%%%%%%%%%%

 Since the usual field-theoretical formalism implements the \emph{annihilation/creation interpretation} of Ps, 
this entails which quantities one seeks to compute and which ones not, and how this is being done. 
 In particular, we note that if one assumes that electron and anti-electron really annihilate each other, it makes no
sense to ask ``In which state are they?'' once they are gone.
 Thus, to compute the energy of each of the two photons which are being emitted in opposite directions (due to momentum conservation)
in the CM frame of Ps, one just has to halve the relativistic energy of the hydrogenic p-Ps ground state.
 The theoretical high-precision computation of this p-Ps ground state is done perturbatively, expressed as a formal series in powers of, 
and powers of the logarithm of, Sommerfeld's fine structure constant $\alpha_{\mbox{\tiny{S}}}^{} = \frac{e^2}{\hbar c} \approx \frac{1}{137.036}$,
which serves as the dimensionless coupling constant between electron and anti-electron.
 A brief recap of this standard procedure follows:

\medskip 
\noindent
I) Ignoring gravity, the vacuum is defined to have energy zero.

\medskip 
\noindent
II) Adding a theoretically ``non-interacting electron and anti-electron pair'' to the vacuum (technically: replacing $\alphaS$ by $0$), 
the lowest energy $E_g(0)$ of such a two-particle ``universe'' is just the sum of their two rest energies, $E_g(0) = 2\mEL c^2$. 

\medskip 
\noindent
III) One next takes their electric Coulomb interaction $-e^2/r$ into account, obtaining an additive correction 
to the ``free ground state'' energy which, to leading order in powers of $\alphaS$, is given by the lowest eigenvalue of 
the hydrogenic Bohr spectrum with $Z=1$ and reduced mass $\mu = \mEL/2$, 
\begin{equation}\label{eq:BOHRspecPs}
E_{\mbox{\tiny{Ps}}}^{\mbox{\tiny{Bohr}}}
 = \left\{- \frac{\mEL c^2 \alpha_{\mbox{\tiny{S}}}^2}{4n^2}\right\}_{n\in\Nset};
\end{equation}
recall that \refeq{eq:BOHRspecPs} can be obtained by semi-classical techniques, or by solving either the two-body Schr\"odinger
or Pauli equation for the electron-positron system with the Coulomb interaction $-e^2/r$ as the only interaction potential, with
center-of-mass coordinates separated off. 
 Thus, to $O(\alphaS^2)$ included the hydrogenic Ps ground state energy is
\begin{equation}
E(\alphaS) = 2\mEL c^2\left(1 - \tfrac{1}{8}\alpha_{\mbox{\tiny{S}}}^2 + O(\alpha_{\mbox{\tiny{S}}}^4)\right).
\end{equation}

\medskip 
\noindent
IV) The $O(\alpha_{\mbox{\tiny{S}}}^4)$ corrections are obtained by adding the quantum mechanical expected values,
taken with respect to the ground state wave function found by solving the two-body Pauli equation in step III (after separating
off the center-of-mass degrees of freedom), of the following operators: the relativistic kinetic energy expanded one order beyond 
the Newtonian term, the Darwin term,
and the spin-spin coupling term --- note that the spin-orbit coupling term yields no contribution in the p-Ps's $\ell=0$-state. 
 This yields (cf. \cite{KarplusKlein})
\begin{equation}\label{eq:EtoORDER4}
E(\alphaS) = 2\mEL c^2\left(1 - \tfrac{1}{8}\alpha_{\mbox{\tiny{S}}}^2 - \tfrac{5}{32} \alpha_{\mbox{\tiny{S}}}^4
+ O(\alpha_{\mbox{\tiny{S}}}^5)\right).
\end{equation}

\medskip 
\noindent
V) The $O(\alpha_{\mbox{\tiny{S}}}^5)$, and $O(\alpha_{\mbox{\tiny{S}}}^5 \ln\frac1\alphaS)$ corrections have also 
been computed in \cite{KarplusKlein}.
 By now the expansion has been pushed to $O(\alpha_{\mbox{\tiny{S}}}^7 \ln^2\frac1\alphaS)$.\footnote{These very 
   high precision calculations are needed to 
   match the very precise experimental measurements of the spectral line of the o-PS to p-Ps hyperfine transition which, because of
   cancellations of the lower order terms in the expansion of the hydrogenic o-Ps and p-Ps ground state energies, releases the
   energy $ \tfrac{7}{12} \alpha_{\mbox{\tiny{S}}}^4\mEL c^2 $, corrected by higher order terms; see, e.g., \cite{positronium,PsS}.}

 For the purpose of computing the energy released through ``annihilation'' of the hydrogenic p-Ps ground state, one does not need 
such high precision calculations.
 Already the terms displayed in \refeq{eq:EtoORDER4} are more than enough for this goal.
 Indeed, assuming that Positronium  will annihilate completely when its hydrogenic p-Ps ground state decays by emitting two photons, 
it follows from energy and momentum conservation that in the CM frame each photon carries away an energy 
$E_\gamma^{}= \mEL c^2\left(1-\tfrac{1}{8}\alphaS^2 - \tfrac{5}{32} \alphaS^4+ O(\alphaS^5)\right)\approx 511$keV; now noting that
$\tfrac{5}{32} \alphaS^4 \approx 4.43 \times 10^{-10}$, it is clear that even if all $O(\alpha_{\mbox{\tiny{S}}}^4)$ terms are ignored 
one still achieves a relative precision of 8 decimal places. 
%\newpage

%%%%%%%%%%%%%%%%%%%%%%%%%%%%%%%%%%%%%%%%%%%%%%%%%%%%%%%%%%%%%%
%%%%%%%%%%%%%%%%%%%%%%%%%%%%%%%%%%%%%%%%%%%%%%%%%%%%%%%%%%%%%%
                \subsection{Critique of the perturbative treatment}\label{sec:pPsCRITIC}
%%%%%%%%%%%%%%%%%%%%%%%%%%%%%%%%%%%%%%%%%%%%%%%%%%%%%%%%%%%%%%
%%%%%%%%%%%%%%%%%%%%%%%%%%%%%%%%%%%%%%%%%%%%%%%%%%%%%%%%%%%%%%

  In the calculation of the Ps pseudo-ground state energy described above, only the electrostatic Coulomb attraction between a point electron 
and a point anti-electron has been handled non-perturbatively,\footnote{Since the $O(\alphaS^2)$ energy term in \refeq{eq:EtoORDER4}
  is just a small correction to the rest energy term $2\mEL c^2$, it has the appearance of being just a perturbation to the rest energy term.
  However, it cannot be computed perturbatively.\vspace{-10pt}}
while all $O(\alpha^4)$ terms, in particular, the magnetic spin-spin interaction (recall that the spin-orbit interaction does not contribute 
in the $\ell=0$ state) have been treated merely perturbatively.
 The conventional rationale for this procedure is the following quasi-classical rule-of-thumb estimate: in the Pauli ground state of 
Ps with purely Coulombic interaction the two particles are most likely a distance $2\frac{\hbar}{\mEL c}\frac1\alphaS$ 
(the ``Bohr radius of Positronium'') apart. 
 At this distance, the electric interaction energy is $-\frac12\alphaS^2\mEL c^2$ while the 
attractive magnetic dipole-dipole interaction energy attributed to the two spins, taking the Bohr magneton $\frac12 \frac{e\hbar}{\mEL c}$
as magnetic dipole strength, is a factor $\alphaS^2$ smaller (ignoring numerical factors of order 1) --- thus a perturbative 
calculation should suffice to account for the influence of spin onto the hydrogenic ground state of Ps. 

 Our point of departure is the difference in the distance scaling laws of the electric monopole-monopole and magnetic dipole-dipole interactions.
 While the electric monopole-monopole interaction energy scales $\propto -1/r$, the magnetic dipole-dipole one scales $\propto -1/r^3$. 
 Of course, the two types of interaction energy terms also have different coupling parameters, but their ratio is dimensionless and 
determines a unique distance at which their ratio equals unity. 
 And so, since a perturbative treatment of the spin-spin coupling is justified \emph{only} if the electric 
pair energy dominates the magnetic one, the validity of perturbative QED calculations is restricted to the far side of the break-even distance,
as the electric monopole-monopole interaction weakens much more slowly than the attractive magnetic dipole-dipole interaction
when the particle distance $r$ increases. 
 Short of the break-even distance, the situation is precisely the other way round! 
 
 The upshot is: there might be magnetically bound states of Positronium with energies way below the hydrogenic ground state energy, whose
computation was outlined in the previous subsection. 
 Their computation requires a non-perturbative treatment of the magnetic interactions of electron and anti-electron.
 Our calculations are non-perturbative and show the viability of a magnetically bound ground state of Ps near zero energy.

%%%%%%%%%%%%%%%%%%%%%%%%%%%%%%%%%%%%%%%%%%%%%%%%%%%%%%%%%%%%%%
%%%%%%%%%%%%%%%%%%%%%%%%%%%%%%%%%%%%%%%%%%%%%%%%%%%%%%%%%%%%%%
                \subsection{\hspace{-5pt}A non-perturbative semi-classical treatment of p-Ps}\label{sec:unPERTURBED}
%%%%%%%%%%%%%%%%%%%%%%%%%%%%%%%%%%%%%%%%%%%%%%%%%%%%%%%%%%%%%%
%%%%%%%%%%%%%%%%%%%%%%%%%%%%%%%%%%%%%%%%%%%%%%%%%%%%%%%%%%%%%%

 A definitive non-perturbative investigation of Positronium cannot be carried out by summing Feynman diagrams, and ``non-perturbative QED''
does not yet exist because the renormalization flow to remove its ultraviolet cutoffs has not been completed; many experts actually believe 
that the non-perturbative removal of the UV cutoffs is \emph{impossible}.
 In such a situation the best one can do is to resort to reasonable approximations, using techniques which in the past have been employed
successfully already.
 For instance, the Bethe--Salpeter equation is such an approximate tool;
interestingly, in \cite{GoldsteinJ} it was argued that a zero-energy bound state of two equal mass nucleons is feasible.

 In this paper, we will work first with Bohr--Sommerfeld type calculations of the energy spectrum.
 We recall the remarkable exact coincidence of the Dirac \emph{Hydrogen energy eigenvalue spectrum} \cite{Darwin}
with Sommerfeld's fine structure formula \cite{fineSTRUKTUR}.
 Both are calculated in the Born--Oppenheimer approximation, i.e. with the proton assumed fixed.
 However, since Dirac's model incorporates electron
spin while Sommerfeld's does not, it is clear that the accounting of {angular momentum} differs in the two models.
  The special circumstances of why nevertheless the two energy spectra agree have been explained in detail in the beautiful work of 
Keppeler \cite{Keppeler}, who analysed the semi-classical limit of the Dirac equation for Hydrogen, and discovered a generalization of 
the Einstein--Brillouin--Keller quantization rules to spinning particle motions.
 The exact coincidence of the spectra is an exception and not the rule, but Keppeler's work explains this exception, 
and also the (smaller) discrepancies in the more general situations.
 Since our goal is not high-precision computations but a qualitative --- yet mathematically compelling --- demonstration that a 
magnetically tightly bound Positronium quantum state is possible, a Bohr--Sommerfeld treatment should be good enough for the purpose.
 We remark that the Bohr--Sommerfeld quantization is not a Lorentz covariant technique; it may be possible to formulate a Lorentz covariant 
two-body version, though (see  \cite{Synge}).

 To get a feeling for the accuracy that can be expected from a Bohr--Sommerfeld calculation of the Positronium energy spectrum, 
we first compute the hydrogenic spectrum without magnetic interactions. 
 Since we are not trying to compute any fine structure, we simply work with the circular Bohr orbits 
rather than Darwin--Sommerfeld rosetta orbits. 

 Then we add the magnetic interactions of two point dipoles and find the first indication for the size of a tightly magnetically 
bound state with $\approx 0$ energy --- note that the Ps Hamiltonian with a magnetic interaction $\propto -1/r^3$ is unbounded 
below (which is the case for the classical as well as for the Pauli Hamiltonian; a Dirac Hamiltonian is of course unbounded below 
even without interactions, but that is a different issue), and the size estimate comes from where the energy function has
its radial zero.

 To avoid the unboundedness below of the Ps Hamiltonian, the only way out is to assume that the magnetic $\propto -1/r^3$ interaction does
not hold all the way down to zero distance but mellows into something less strongly diverging.
 One way to accomplish this is to assume that electron and anti-electron are not true points but have a nontrivial geometric structure.
 We follow Max Born (see \cite{Rao}) and assume the particles are tiny rings\footnote{We could also call them ``closed strings'' or ``loops,''
 although ``Born's rings'' are not the strings of string theory.
 Interestingly, 
  in the zero-gravity limit of the general-relativistic Kerr--Newman spacetime the source of its static electromagnetic fields is also a 
  spacelike ring, see \cite{TZzGKN}. 
   Different from Born's ring, the z$G$KN ring is a topological defect of a two-sheeted electromagnetic spacetime, and its charge and current
have a curious sesquipole structure, yet far from the ring the electric field has monopole and the magnetic field dipole asymptotics.
   The Dirac operator of a z$G$KN ring interacting with a point charge has been studied in \cite{KTZzGKNa,KTZzGKNb}, and a 
study for the Dirac spectrum of two z$G$KN type rings with Ps parameters is underway.\vspace{-5pt}}
which carry charge and current.
 However, different from Born's suggestion that the electromagnetic fields be computed using the Born--Infeld
modification of the Maxwell--Lorentz equations (which nobody has been able to do for ring sources so far), we are using, first, 
the conventional Maxwell--Lorentz equations (but then have to subtract the infinite self-energies), and second, 
the Bopp \cite{BoppA,BoppB}, Land\'e--Thomas \cite{landethomas}, Podolsky \cite{Podolsky} modification of the Maxwell--Lorentz 
field equations, used also by Feynman as UV cutoff \cite{FeynmanBOPP}.
 It yields finite self-energies and a finite magnetic flux through a ring. 

 A very subtle issue is the size of the magnetic moment assigned to the ring particles. 
 It would seem obvious that the magnitude of the magnetic moment should be the Bohr magneton. 
 However,  \emph{the Bohr magneton} is associated with the electron spin
which, as emphasized by Pauli, is implemented in the Pauli and Dirac equations for point particles through the Pauli $\sigma$ matrices 
acting on the spinors, and so \emph{has nothing to do with any electron structure}.
 Any magnetic moment associated with a structure of the electron therefore has to come \emph{in addition} to the Bohr magneton.
 As argued recently also in \cite{KTZzGKNb}, it is very suggestive to equip the ring structure of Born's electron 
with the electron's anomalous magnetic moment\footnote{Note that this leading order term of $\mu_{\mbox{\tiny{anom}}}$, 
  though computed perturbatively using QED, is independent of $\hbar$, and may thus
  be associated with a ``classical structure'' of the electron.
  It is of course also logically possible that any electron structure makes yet another contribution to the magnetic moment of the electron, 
   in addition to the Bohr magneton size moment in the spinor structure of Pauli and Dirac equations, and in addition to the 
   anomalous magnetic moment.}
$\mu_{\mbox{\tiny{anom}}}\approx \frac{1}{4\pi}\frac{e^3}{mc^2}$.
 In the final section we will comment on other choices.

 To keep the formulas simpler, we will from now on use the following units: the energy unit is $\mEL c^2$, and that of mass is $\mEL$; 
the unit of momentum is $\mEL c$; the unit of angular momentum is $\hbar$; the unit of length is $\hbar/\mEL c$; the unit of 
electric charge is $e$.

%%%%%%%%%%%%%%%%%%%%%%%%%%%%%%%%%%%%%%%%%%%%%%%%%%%%%%%%%%%%%%
                \subsubsection{Two point particles with electric charges}\label{sec:SEMIclassCOULOMB}
%%%%%%%%%%%%%%%%%%%%%%%%%%%%%%%%%%%%%%%%%%%%%%%%%%%%%%%%%%%%%%
 In the CM frame of a mutually circulating electron and anti-electron, and expressed in the units just stipulated,
their classical energy function reads
\begin{equation}\label{eq:HamFCTcircCOULOMB}
 H(r,p) :=  % 2 \sqrt{\mEL^2 c^4 + p^2 c^2} - e^2\tfrac1r,
              2 \sqrt{1 + p^2 } - \tfrac{\alphaS}{r},
\end{equation}
with $p$ and $r$ constants.
 To find the principal Bohr--Sommerfeld energy values of Positronium one needs to minimize \refeq{eq:HamFCTcircCOULOMB}
subjected to Bohr's quantization condition $pr=n$ %$=n\hbar$
 with $n=1$; here, $p$ is the magnitude of the electron's or anti-electron's momentum, 
both assumed to move in a circular path around a common center, $r$ is their distance, and $n\in\Nset$ is Bohr's orbital angular quantum 
number.\footnote{Since 1925 we know that $n$ is not the correct quantum-mechanical orbital angular momentum quantum number, but 
  this is immaterial for the computation of the energy levels.\vspace{-5pt}} 
 Thus, to compute the principal Bohr--Sommerfeld energy values of Positronium one needs to minimize 
\begin{equation}
V_n^{(1)}(r) := 
                   2 \sqrt{1 +  \tfrac{n^2}{r^2}} - \tfrac{\alphaS}{r}
\end{equation}
with respect to $r$. 
 (See Fig.~1 for $V_n^{(1)}(r)$ when $n=1$ (green) and $n=2$ (blue).)

 Minimization of $V_n^{(1)}(r)$, for any $n\in\Nset$, yields the energy spectrum
\begin{equation}\label{eq:BOHRspecPsAGAIN}
E_n(\alphaS) =  2 \sqrt{1 - \tfrac{\alphaS^2}{4n^2}}.
\end{equation}
 Maclaurin expansion in powers of $\alphaS$ yields
\begin{equation}\label{eq:BOHRspecPsTAYLOR}
E_n(\alphaS) = 2  \left(1 - \tfrac{1}{8n^2}\alpha_{\mbox{\tiny{S}}}^2 - 
\tfrac{1}{128n^4} \alpha_{\mbox{\tiny{S}}}^4 + O(\alpha_{\mbox{\tiny{S}}}^6)\right).
\end{equation}
 The ground state energy of Coulombic Ps using Bohr--Sommerfeld rules is obtained by setting $n=1$ in \refeq{eq:BOHRspecPs}.
 After subtracting the spin-spin interactions, the $O(\alpha_{\mbox{\tiny{S}}}^4)$ term in \refeq{eq:BOHRspecPsTAYLOR} differs 
from the one in \refeq{eq:EtoORDER4} by a factor 4; so this calculation reproduces \refeq{eq:EtoORDER4} up to terms of 
$O(\alpha_{\mbox{\tiny{S}}}^2)$ included but only gives the sign of the $O(\alpha_{\mbox{\tiny{S}}}^4)$ term correctly.\vspace{-10pt}
 
%%%%%%%%%%%%%%%%%%%%%%%%%%%%%%%%%%%%%%%%%%%%%%%%%%%%%%%%%%%%%%
                \subsubsection{Two point particles with electric charges and magnetic dipoles}\label{sec:SEMIclassCOULOMBandAMPERE}\vspace{-5pt}
%%%%%%%%%%%%%%%%%%%%%%%%%%%%%%%%%%%%%%%%%%%%%%%%%%%%%%%%%%%%%%

 The anomalous magnetic moment of the particles, itself computed perturbatively from QED, 
yields a correction to the spin-orbit coupling already inherent in Dirac's equation, plus a magnetic dipole-dipole type interaction
energy which is itself treated with first-order perturbation theory, each yielding a tiny correction to the Coulombic spectrum. 
 Here we add an attractive anomalous magnetic dipole-dipole interaction in an ad-hoc manner to \refeq{eq:HamFCTcircCOULOMB}, implement 
Bohr's quantization condition, and look for (local) minima of
\begin{equation}
V_n^{(2)}(r) := 
  2 \sqrt{1 +  \tfrac{n^2}{r^2}} - \tfrac{\alphaS}{r} - \tfrac{1}{8\pi^2}\tfrac{\alphaS^3}{r^3}
\end{equation}
with respect to $r$, for $n=1$, see Figs.~1 and 2.\vspace{-1.5truecm} 
 
\begin{figure}[H]
\includegraphics[width=12.25cm]{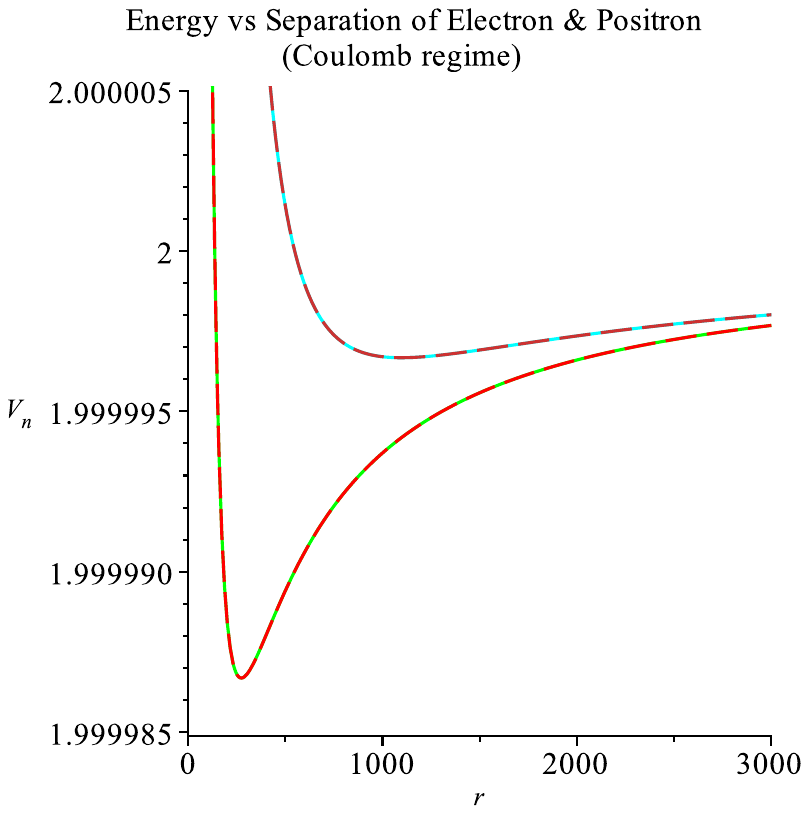} \label{fig:HYDROGENICregime}
\vspace{-7.5truecm}
\end{figure}
\centerline{\hspace{-4truecm} Figure 1}

 Fig.~1 shows the usual hydrogenic regime which is dominated by the balancing of the kinetic energy and the
Coulomb interaction energy of the two point charges.
 Both $V_n^{(1)}(r)$ and $V_n^{(2)}(r)$ are plotted versus $r$, each for $n=1$ (lower curves) and $n=2$ (upper curves). 
 Recall that the unit for $V_n$ is $\mEL c^2$, the unit for $r$ is $\frac{\hbar}{\mEL c}$.
 The local minimum for $n=1$ occurs for $r$ roughly equal to the ``Bohr radius of Positronium,''
 $2 \frac{\hbar}{\mEL c}\frac1\alphaS$, i.e. $r\approx 274$ in our electron Compton length units.
 For each $n$ value the two pertinent curves are virtually indistinguishable at this leading-order energy scale, 
confirming that a perturbative treatment of magnetic effects is justified in this regime. 

 Not visible on the scale shown in Fig.~1 is the behavior  of both $V_n^{(1)}(r)$ and $V_n^{(2)}(r)$ for separations $r$ 
very much smaller than the ``Bohr radius of Positronium.''
 This very different scale is shown in Fig.~2, now only for $n=1$; the color code is the same as in Fig.~1.
\vspace{-1truecm}

\begin{figure}[H]
\includegraphics[width=12.25cm]{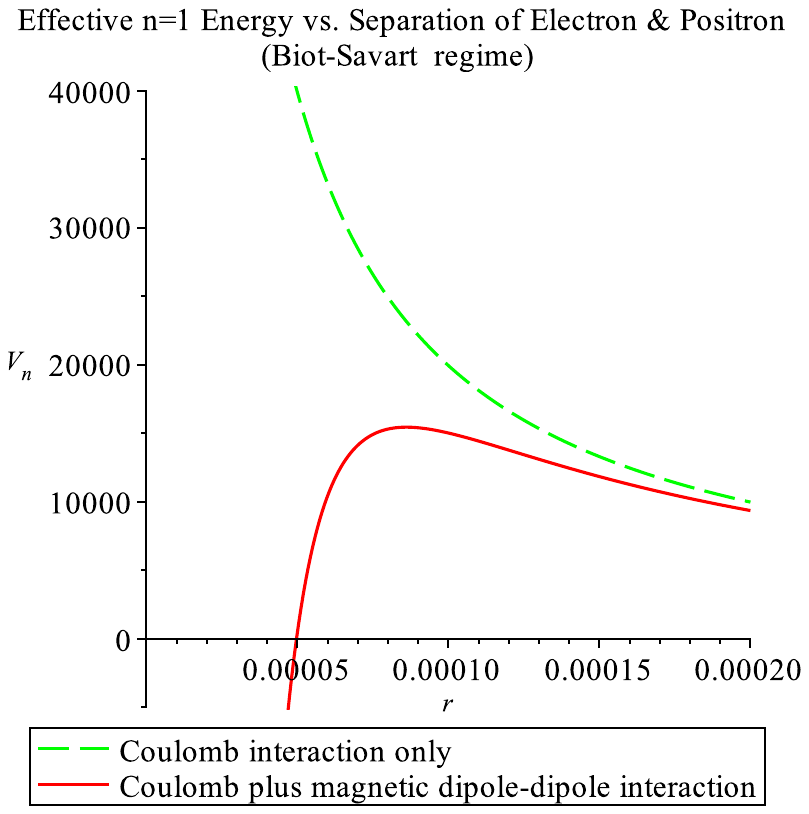} \label{fig:BIOTSAVARTregime}
\vspace{-7truecm}
\end{figure}
\centerline{\hspace{-4truecm} Figure 2}

 Figure 2 reveals that a tightly bound state due to the magnetic dipole-dipole type attraction of electron and positron is
feasible, provided at very short distances the $-1/r^3$ behavior is mollified into something gentler, diverging to $-\infty$ not
faster than $-1/r$ because the relativistic angular momentum barrier scales $\propto 1/r$ for small $r$. 
 The transition from $-1/r^3$ scaling to something milder has to happen for distances smaller than $10^{-4}$ 
reduced Compton lengths of the electron.
 This is accomplished by assuming the electron (and the positron) are little rings, an idea which goes back to Max Born it seems.

%%%%%%%%%%%%%%%%%%%%%%%%%%%%%%%%%%%%%%%%%%%%%%%%%%%%%%%%%%%%%%
    \subsubsection{Two Born rings with electric charges and currents: Part I}\label{sec:SEMIclassCOULOMBandAMPEREringsML}
%%%%%%%%%%%%%%%%%%%%%%%%%%%%%%%%%%%%%%%%%%%%%%%%%%%%%%%%%%%%%%

 We now assume that both electron and positron are tiny rings of radius $R$, carrying an electric charge $\mp e$, and an electric current
$\mp I$ such that $\pi R^2 I\frac1c = \frac{\alphaS}{2\pi} \frac{e\hbar}{2\mEL c} = \frac{1}{4\pi}\frac{e^3}{mc^2}$. 
 Eliminating $I$ this way leaves one with $R$ free to adjust. 
 Given the distance $r$ between the geometrical centers of the two rings, we minimize their mutual electromagnetic energy as computed
with the Maxwell--Lorentz fields.
 The minimum occurs if the rings are co-planar, oriented such that their interaction energy increases with $r$.
 These energies have been computed many times in the literature. 
 With the help of the complete elliptic integrals $\bK$ and $\bE$ (see \cite{AS})
we thus obtain for the to-be-minimized energy function $V_n^{(3)}(r;R)$:\vspace{-5pt}
\begin{alignat}{1}\label{VnML}
V_n^{(3)}(r;R) 
:= 2\sqrt{1+\tfrac{n^2}{r^2}} +  U_{R}(r),
\end{alignat}
where
\begin{alignat}{1}\label{eq:U}
 U_{R}(r)
: = 
& - \tfrac{1}{\pi}\tfrac{\alphaS}{R}\tfrac{1}{\sqrt{1+\tfrac{r^2}{4R^2}}}\bK\biggl(\tfrac{1}{\sqrt{1+\tfrac{r^2}{4R^2}}}\biggr)
\\ \notag 
&-\tfrac{1}{4\pi^3}\tfrac{\alphaS^3}{R^3}
\sqrt{1+ \tfrac{r^2}{4R^2}}\biggl[\biggl(2-\tfrac{1}{1+\tfrac{r^2}{4R^2}}\biggr) {\bK}\biggl(\tfrac{1}{\sqrt{1+\tfrac{r^2}{4R^2}}}\biggr)
-2\bE\biggl(\tfrac{1}{\sqrt{1+\tfrac{r^2}{4R^2}}}\biggr)\biggr]
% U_{\varkappa,R}(r) := - \tfrac{\alphaS}{2\pi R}
%\int_0^{\pi} \left(1+
%\left(\tfrac{\alphaS}{2\pi R}\right)^2
%\cos(2\varphi) \right)
%\tfrac{1-\exp\bigl(-2\varkappa R \sqrt{\sin^2(\varphi)+\frac{r^2}{4R^2}}\bigr)}{\sqrt{\sin^2(\varphi)+\frac{r^2}{4R^2}}}\drm\varphi,
\end{alignat}

\begin{figure}[H]
\includegraphics[width=12.25cm]{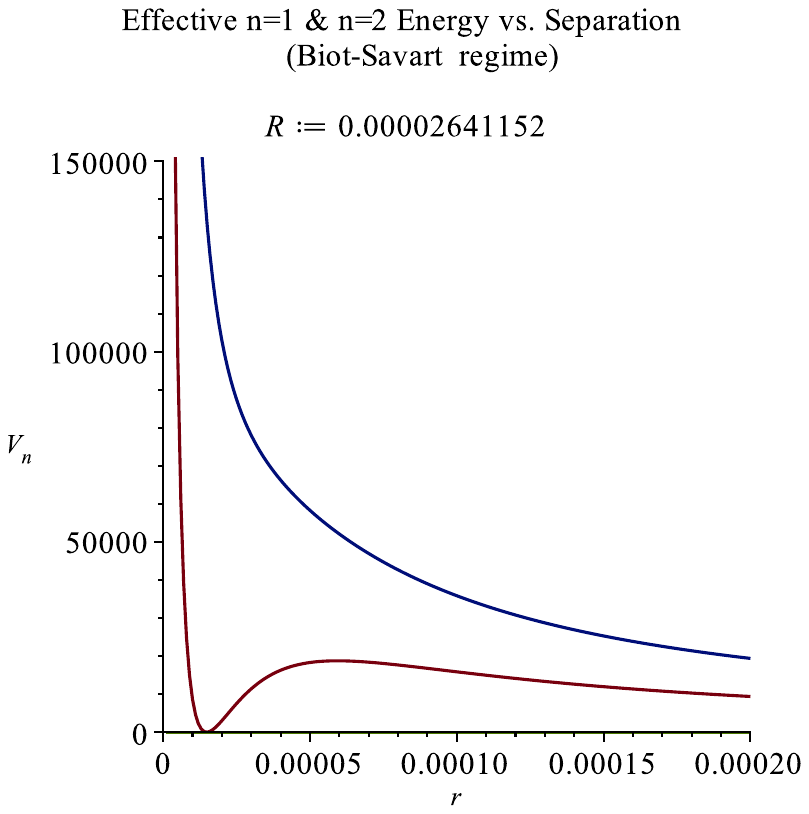} \label{fig:BIOTSAVARTregimeMLrings}
\vspace{-7.5truecm}
\end{figure}
\centerline{\hspace{-4truecm} Figure 3}
% R = 0.26411517199 10^{-4} = 0.000026411517199

 In Fig.~3 we show $V_n^{(3)}(r;R)$ for $n=1$ (lower curve) and $n=2$ (upper curve) in the Biot--Savart dominated regime when\footnote{The
 many decimal places are needed because the ground state energy is given by the difference of two huge numbers. 
 For instance, dropping the digit ``5'' in the last decimal place of the numerical factor of $\alphaS^2$ results in a negative ground state energy.}
$R=0.49597832375\alphaS^2$.

 We see a global minimum of the $n=1$ curve for $r\approx 1.3 \times 10^{-5}$ reduced electron Compton lengths that corresponds to a tightly bound 
$n=1$ state with energy $\approx 0$, indeed; we also see that for $n=2$ there is no additional tightly bound state (in fact, this is easily
shown to be true for any $n>1$).
 We also remark that in the Coulomb dominated regime (not shown) the energy curves are virtually indistinguishable from those shown in Fig.~1, 
producing the hydrogenic Bohr spectrum, as they should.

%%%%%%%%%%%%%%%%%%%%%%%%%%%%%%%%%%%%%%%%%%%%%%%%%%%%%%%%%%%%%%
   \subsubsection{Two Born rings with electric charges and currents: Part II}\label{sec:SEMIclassCOULOMBandAMPEREringsMBLTP}
%%%%%%%%%%%%%%%%%%%%%%%%%%%%%%%%%%%%%%%%%%%%%%%%%%%%%%%%%%%%%%

 We continue to assume that both electron and positron are tiny rings of radius $R$, carrying an electric charge $\mp e$, and an electric current
$\mp I$ such that $\pi R^2 I\frac1c = \frac{\alphaS}{2\pi} \frac{e\hbar}{2\mEL c}= \frac{1}{4\pi}\frac{e^3}{mc^2}$.  
 As before, eliminating $I$ this way leaves one with $R$ free to adjust. 

 However, to demonstrate the robustness of a tightly magnetically bound Ps state, we now compute the interaction energies with 
the Bopp--Land\'e--Thomas--Podolsky modification of the Maxwell--Lorentz field equations, 
see \cite{BoppA,BoppB}, \cite{landethomas}, \cite{Podolsky} for the original works, \cite{FeynmanBOPP} for a follow-up study, 
and \cite{KTZonBLTP} for a modern rigorous assessment of its classical merits.
 In the MBLTP field theory the $\HV$ and $\DV$ fields are the same as in the Maxwell--Lorentz theory, but
the law of the electromagnetic vacuum of the ML theory, $\HV=\BV$ and $\DV=\EV$, is replaced by 
partial differential equations which in the static limit reduce to
\begin{alignat}{1}
        \HV(\sV)  
&= \label{eq:BLTPlawBandH}
       \left(1  - \varkappa^{-2}\Delta\,\right) \BV(\sV) \, ,
\\
        \DV(\sV) 
&=
        \left(1  - \varkappa^{-2}\Delta\,\right) \EV(\sV) \, ;
\label{eq:BLTPlawEandD}
\end{alignat}
here, $\Delta$ is the Laplacian, and $\varkappa$ is Bopp's fundamental inverse length 
(which is undetermined as of yet; however, in \cite{CKP} the Schr\"odinger spectrum of Hydrogen is studied and it is found
that the currently available precision of the Lyman $\alpha$ fine structure suggests a lower limit of $\approx 10^{18}$m$^{-1}$
for $\varkappa$, or $O(10^5)$ reciprocal reduced Compton length units.).

 While the solutions of the Maxwell--Lorentz field equations for a point or ring electron have an infinite total field energy,
the pertinent solutions of the Maxwell--Bopp--Land\'e--Thomas--Podolsky field equations have a finite field energy\footnote{Unfortunately, 
  the field energy of point dipoles with a BLTP modification of the Maxwell--Lorentz field equations is still infinite.}
 --- a welcome feature of this theory!
 The interaction energy between two ring particles carrying electric charges and currents is now defined as usual as 
the total field energy of such a configuration minus the self-field energies of its constituents, but unlike for Maxwell--Lorentz fields,
this does not now amount to an uncomfortable ``infinite self-energy subtraction.''
 In the static limit (relevant to computing the interaction energy needed for the determination of quantum mechanical energy spectra)
the total electromagnetic energy of the MBLTP fields is given by the familiar expression\footnote{The non-negativity
   of the integral can be shown with the help of the BLTP vacuum law, \refeq{eq:BLTPlawBandH} expressing $\HV$ in terms of $\BV$, 
   and \refeq{eq:BLTPlawEandD} expressing $\DV$ in terms of $\EV$, followed by integration by parts.
    We remark that \refeq{eq:MBLTPfieldENERGY} is valid only in static situations.}
\begin{equation}\label{eq:MBLTPfieldENERGY}
E_{\mbox{\tiny{field}}}  = \frac{1}{8\pi}\int_{\Rset^3}\left({\EV\cdot\DV}+ {\BV\cdot\HV}\right) d^3s \geq 0.
\end{equation}
 For the ring particles we are discussing right now this is easily reduced to one-dimensional quadratures which, to the best of our
knowledge, have not been expressed in terms of already known special functions (such as complete elliptic integrals in the 
Maxwell--Lorentz case, for instance).
 Thus we now have to minimize 
\begin{alignat}{1}\label{VnMBLTP}
V_n^{(4)}(r;R,\varkappa) 
%:= 2\sqrt{1+\tfrac{n^2}{r^2}} & -
%\tfrac{\alphaS}{2\pi}
%\int_0^{2\pi} \tfrac{1-\exp\left(-\varkappa\sqrt{2R^2+r^2-2R^2\cos(\varphi)}\right)}{\sqrt{2R^2+r^2-2R^2\cos(\varphi)}}\drm\varphi
%\\ \notag
%& -
%\left(\tfrac{\alphaS}{2\pi}\right)^3\tfrac{1}{R^2}
%\int_0^{2\pi} \cos(\varphi) \tfrac{1-\exp\left(-\varkappa\sqrt{2R^2+r^2-2R^2\cos(\varphi)}\right)}{\sqrt{2R^2+r^2-2R^2\cos(\varphi)}}\drm\varphi \\
:= 2\sqrt{1+\tfrac{n^2}{r^2}} & -
\tfrac{\alphaS}{2\pi R}
\int_0^{\pi} \tfrac{1-\exp\bigl(- 2\varkappa R \sqrt{\sin^2(\varphi)+\frac{r^2}{4R^2} }\bigr)}{\sqrt{\sin^2(\varphi)+\frac{r^2}{4R^2}}}\drm\varphi
\\ \notag
& -
\left(\tfrac{\alphaS}{2\pi R}\right)^3 
\int_0^{\pi} \cos(2\varphi) \tfrac{1-\exp\bigl(-2\varkappa R \sqrt{\sin^2(\varphi)+ \frac{r^2}{4R^2}}\bigr)}{\sqrt{\sin^2(\varphi)+\frac{r^2}{4R^2}}}\drm\varphi
\end{alignat}
w.r.t. $r$, and with $R$ and $\varkappa$ to be determined wisely --- by this we mean the following.

 Since the BLTP vacuum law of electromagnetism not only renders the field energies etc. finite, but also the magnetic flux through
a ring particle which carries an electric charge and current (neither is true for the fields computed with the ML vacuum law),
 we impose a heuristic relationship on $\varkappa$ and $R$ by demanding that the magnetic self-field flux through a ring equals
the empircal magnetic flux quantum, $e\frac1\alphaS$. 
 This yields the constraint
\begin{equation}\label{eq:magnFLUXconstraint}
R = \tfrac{\alphaS^2}{2\pi}\int_0^\pi \cos(2\varphi)\tfrac{1-\exp\bigl(-2\varkappa R\sin(\varphi)\bigr)}{\sin(\varphi)}\drm\varphi.
\end{equation}
 With \refeq{eq:magnFLUXconstraint} in place we have adjusted the product $\varkappa R$ until the lowest energy minimum was essentially zero.
 This has yielded $R\approx 2.57\times 10^{-5}$ and $\varkappa \approx 1.8\times 10^5$.

 In Fig.~4 we show $V_n^{(4)}(r;R,\varkappa)$ (solid lines) for $\varkappa \approx 1.8\times 10^5$ and $R\approx 2.57\times 10^{-5}$, both for $n=1$ 
and $n=2$, together with a truncated version (dashed lines) of $V_n^{(4)}(r;R,\varkappa)$ which omits the magnetic interactions.
  The figure shows precisely one bound state in the Biot--Savart regime, associated with the $V_1^{(4)}(r;R,\varkappa)$ curve, while the
$V_n^{(4)}(r;R,\varkappa)$ curves for $n>1$ do not have a local minimum in the Biot--Savart regime --- they all do have a global minimum
in the Coulomb regime, of course, where the $n=1$ curve has a local minimum in addition to the global one shown in Fig.~4.
 As before, the continuation of the energy curves to the Coulomb dominated regime (not shown) reveals that there they 
 are virtually indistinguishable from those shown in Fig.~1, producing the hydrogenic Bohr spectrum, as they should.

 Although the fixing of $\varkappa$ and $R$ is admittedly heuristical, it is perfectly reasonable. 
 What is comforting is that the obtained value of $\varkappa$ is in line with the lower estimate for it obtained in \cite{CKP},
and that the $r$ value of the global energy minimum comes out roughly the same in all three calculations presented here. 
 Moreover, the value of the ring radius $R$ comes out about the same in the two ring model calculations presented here. 
 Thus our semi-classical calculations 
 demonstrate that the existence of a magnetically bound ground state of Positronium with energy near zero is a viable conjecture.\vspace{-10pt}

\begin{figure}[H]
\includegraphics[width=12.25cm]{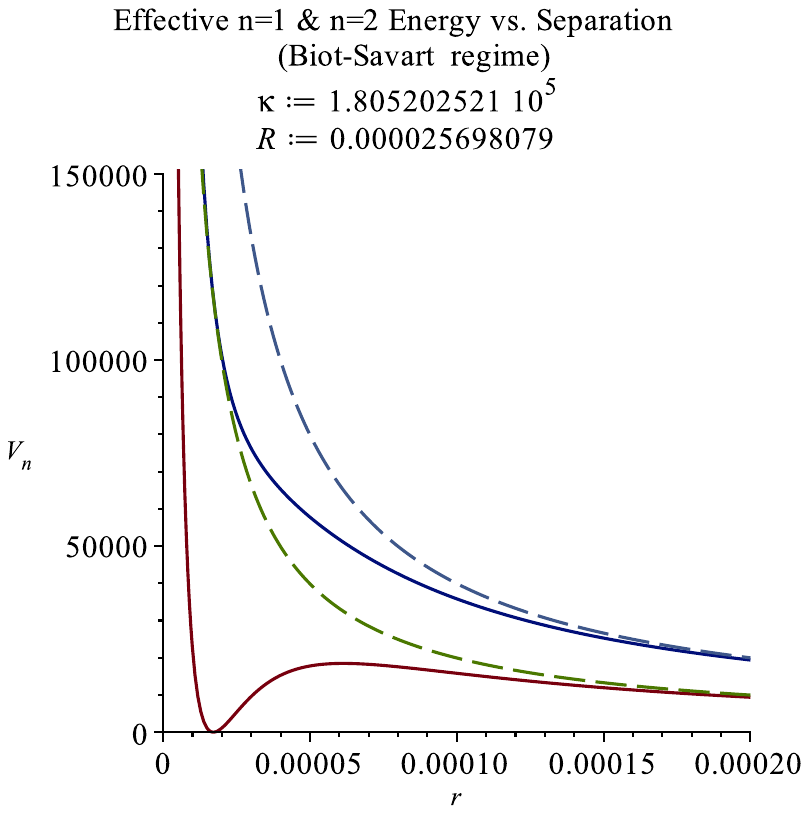} \label{fig:GROUNDstateMBLTP}
\vspace{-7.5truecm}
\end{figure}
\centerline{\hspace{-4truecm} Figure 4}

 In Fig.~5 we show a blow-up (or zoom-in) of the $V_1^{(4)}(r;R,\varkappa)$ curve, with
$\varkappa \approx 1.8\times 10^5$ and $R\approx 2.57\times 10^{-5}$. 
  It is obvious that the ground state energy is practically zero.
\vspace{-1truecm}
\begin{figure}[H]
\includegraphics[width=12.25cm]{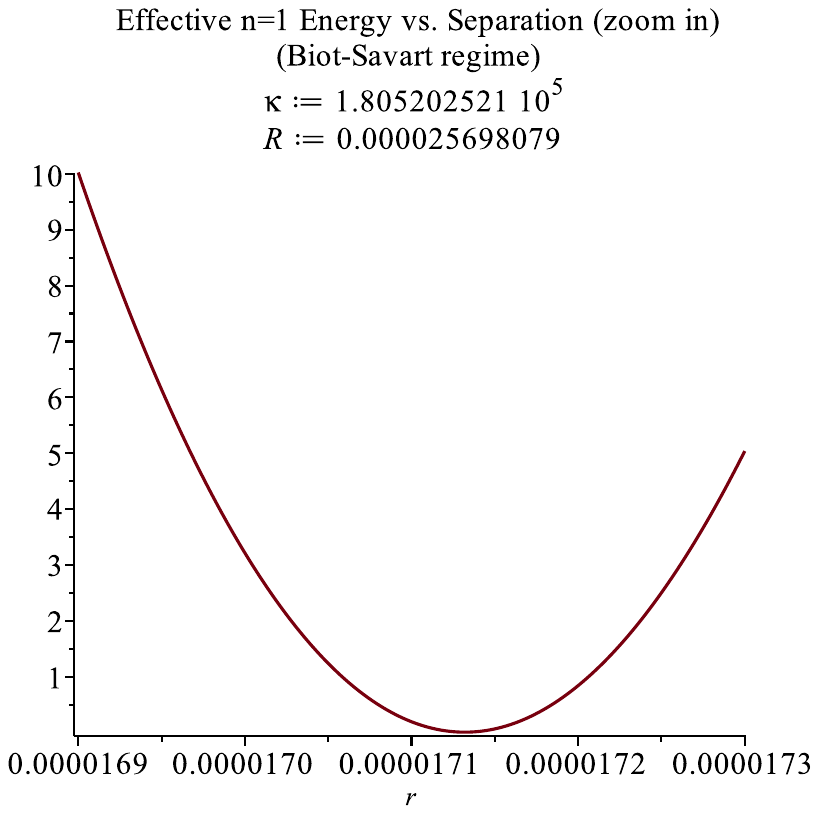} 
\vspace{-7.5truecm}
\end{figure}
\centerline{\hspace{-4truecm} Figure 5}

%%%%%%%%%%%%%%%%%%%%%%%%%%%%%%%%%%%%%%%%%%%%%%%%%%%%%%%%%%%%%%
%%%%%%%%%%%%%%%%%%%%%%%%%%%%%%%%%%%%%%%%%%%%%%%%%%%%%%%%%%%%%%
                \subsection{A non-perturbative quantum-mechanical model of p-Ps}\label{sec:unPERTURBEDsqrtKG}
%%%%%%%%%%%%%%%%%%%%%%%%%%%%%%%%%%%%%%%%%%%%%%%%%%%%%%%%%%%%%%
%%%%%%%%%%%%%%%%%%%%%%%%%%%%%%%%%%%%%%%%%%%%%%%%%%%%%%%%%%%%%%

 In this subsection we demonstrate that the low-lying ground state energy of Ps we found in subsections \ref{sec:SEMIclassCOULOMBandAMPEREringsML} 
and \ref{sec:SEMIclassCOULOMBandAMPEREringsMBLTP} is not just a fluke of the Bohr--Sommerfeld technique but corresponds to a low-lying ground state 
eigenvalue of the following pseudo-differential operator (without spin), given by \refeq{VnML} with $\sqrt{1+\frac{n^2}{r^2}}$ replaced by a
``square root Klein--Gordon'' operator (cf. \cite{laemmerzahl}), which leads to
\begin{alignat}{1}\label{eq:sqrtKGop}
{H} := 2\sqrt{1-\Delta} + U_{R}(r) % U_{\varkappa,R}(r)
\end{alignat}
with $R$ roughly as for the calculations shown in Fig.~3.
% $\varkappa$ and $R$ roughly as for the calculations shown in Fig.~4; we may also drop \refeq{eq:magnFLUXconstraint} and let $\varkappa\to\infty$,
% leading to \refeq{VnML} with $\sqrt{1+\frac{n^2}{r^2}}$ replaced by $\sqrt{1-\Delta}$ (and 
 To be sure, this is not a properly relativistic quantum-mechanical Hamiltonian.
 However, such square root Klein--Gordon operators have been shown to capture some important relativistic features qualitatively correctly,
for instance the Chandrasekhar model of white dwarf stars, obtained from a many-body square root Klein--Gordon operator with Newtonian
gravity in \cite{LiebYau}.

 The potential pair energy $U_R(r)$ is negative, with a $-\ln\frac1r$ singularity as $r\downarrow 0$ and an asymptotic $-\frac1r$ decay as
$r\uparrow\infty$. 
 It follows that $H$ is essentially self-adjoint on 
$C^\infty_c(\Rset^3\backslash\{0\})$ (note that $\sqrt{1-\Delta}$ acts as multiplication operator $\sqrt{1+4\pi^2|\kV|^2}$ in Fourier space,
with the Fourier transform defined as $\widehat{\psi}(\kV):= \int_{\Rset^3}e^{-2\pi i \kV\cdot\qV} \psi(\qV)d^3q$). 
 It also follows that it has infinitely many bound states below the continuum which starts at $2$.
 With $R$ chosen as stipulated, the point spectrum of $H$ to order $\alphaS^2$ features the hydrogenic Bohr spectrum of Ps,
but it also features a ground state eigenvalue $E_g$ way below this Coulombic part of its spectrum.
  The existence of this low-lying $E_g$ can be shown with the help of the Rayleigh--Ritz variational principle, 
$E_g:= \inf_{\psi} \langle \psi|H|\psi\rangle$, with the infimum taken over normalized $L^2$ functions whose Fourier transforms 
are square integrable against the weight $|\kV|$. 
 Any one-parameter family of trial functions $\psi_a(\qV)$ yields an upper bound,
\begin{equation}\label{eq:upperBOUNDonEg}\hspace{-20pt}
E_g \leq \inf_a \langle \psi_a|H|\psi_a\rangle 
= \inf_a \left\{2\!\! \int_{\Rset^3}\!\!\sqrt{1+4\pi^2|\kV|^2}|\widehat{\psi}_a(\kV)|^2 d^3k
+ \int_{\Rset^3}\! U_{R}(r)|\psi_a(\qV)|^2d^3q\!\right\}\!\!.\hspace{-10pt}
\end{equation}
 Choosing the hydrogenic Ps ground state of the non-relativistic Schr\"odinger equation, 
$\psi_a(\qV) := \frac{1}{\sqrt{\pi a^3}} e^{-r/a}$, where $r :=|\qV|$, so that
$\widehat{\psi}_a(\kV):= \frac{8\sqrt{\pi a^3}}{(1+4\pi^2a^2|\kV|^2)^2}$, we find for $R=2.661639\times 10^{-5}$ that
the expected value $\langle\psi_a|H|\psi_a\rangle$ as a function of $a$ qualitatively resembles the graph of $V_1^{(3)}(r;R)$ as function of $r$. 
 In particular, for $a\approx 2/\alphaS$, i.e. near the Bohr radius of Ps, the graph very closely resembles the lower curve in Fig.~2, 
having a local minimum value equal to $2 - \tfrac{1}{4n^2}\alpha_{\mbox{\tiny{S}}}^2+ O(\alphaS^4)$.
 Yet in the neighborhood of $a\approx 1.5726\times 10^{-5}$, the graph resembles the lower curve in Fig.~3; in fact it has a near zero
energy minimum at $a\approx 1.5726\times 10^{-5}$, which gives the upper bound $E_g \leq 0.0535$ for this $R$ value. 
Of course, while the bulk of the hydrogenic spectrum of Ps is sufficiently accurately reproduced by the point spectrum of $H$ 
(though of course not to the same high precision as reviewed in section 2), we have not attempted to accurately 
determine the ground state energy as function of $R$.
 We expect, though, that our variational upper bound $E_g \leq \inf_a \langle \psi_a|H|\psi_a\rangle$ is not too far away from the actual
ground state energy value. 
 In any event, our bound does prove the existence of a low-lying ground state energy of $H$ given in (\ref{eq:sqrtKGop}) when $R$ is roughly
equal to the $R$ value found with the Bohr--Sommerfeld calculation,  as announced.\vspace{-10pt}

%%%%%%%%%%%%%%%%%%%%%%%%%%%%%%%%%%%%%%%%%%%%%%%%%%%%%%%%%%%%%%
%%%%%%%%%%%%%%%%%%%%%%%%%%%%%%%%%%%%%%%%%%%%%%%%%%%%%%%%%%%%%%
%%%%%%%%%%%%%%%%%%%%%%%%%%%%%%%%%%%%%%%%%%%%%%%%%%%%%%%%%%%%%%
                \section{Summary and Outlook}\label{sec:TheEND}\vspace{-5pt}
%%%%%%%%%%%%%%%%%%%%%%%%%%%%%%%%%%%%%%%%%%%%%%%%%%%%%%%%%%%%%%
%%%%%%%%%%%%%%%%%%%%%%%%%%%%%%%%%%%%%%%%%%%%%%%%%%%%%%%%%%%%%%
%%%%%%%%%%%%%%%%%%%%%%%%%%%%%%%%%%%%%%%%%%%%%%%%%%%%%%%%%%%%%%

 In this paper we have investigated the energy spectrum of an electron-positron system bound not only by their electrostatic
Coulomb interactions associated with their charges, but also by their magnetostatic Biot--Savart interactions associated with 
their magnetic moments. 
 Because of their simplicity we first used the Bohr--Sommerfeld quantization techniques which appeared at the dawn of quantum 
physics. 
 While this may seem like a pre-enlightenment approach to quantum physics, one should not forget that Sommerfeld's
fine structure spectrum exactly anticipated the Dirac point spectrum, so that the here computed Positronium spectrum may 
not be too far from the truth either. 
 We found that the magnetic dipole-dipole type $-\frac{1}{r^3}$ attraction (mollified for $r\downarrow 0$), which 
contributes only a weak
perturbation to the hydrogenic part of the spectrum but dominates the Coulomb $-\frac1r$ interaction at shorter distances, causes 
a true ground state of the two-particle system way below the hydrogenic pseudo-ground state energy. 
 We reproduced this result about the Ps spectrum with a variational study of a pertinent square root Klein--Gordon operator. 
 There is no doubt in the author's mind that truly relativistic quantum-mechanical calculations will also confirm the existence 
of a low energy eigenvalue of a Dirac operator suited for a non-perturbative treatment of Ps with Coulomb and Biot--Savart 
interactions.

 By the usual quantum mechanical principles, a transition of Ps from its hydrogenic (i.e. Coulomb dominated) 
p-Ps pseudo-ground state to the low-lying magnetic (i.e. Biot--Savart dominated) ground state level
would be accompanied by the emission of two high-energy photons (two because of angular momentum conservation). 
 Only one such process is known, namely the emission of two 511 keV photons in what is normally interpreteted as the mutual 
annihilation of electron and anti-electron in Ps.
 This means that the low-lying ground state must have an energy near zero, which we were able to arrange
by adjusting the size of the mollifier (here a Born ring) to about $10^{-5}$ reduced electron Compton lengths;
 we emphasize that the size of such a small mollification is compatible with 
the empirical bound on the size of the electron \cite{BrDrPRD}.
 The electromagnetic radiation of a quantum-mechanical transition from a hydrogenic level to 
a near-zero energy magnetic ground state level would be virtually indistinguishable from the radiation due to an actual 
annihilation of Ps.

 In our calculations we used the anomalous magnetic moment of the electron to compute the coupling constant of the Biot--Savart 
type interaction.
 We already remarked in a footnote that the independence from $\hbar$ of the leading order term in the QED expansion 
of $\mu_{\mbox{\tiny{anom}}}$ suggests that it may be associated with a ``classical structure'' of the
electron, yet it is entirely conceivable that this is a coincidence and $\mu_{\mbox{\tiny{anom}}}$ is a purely radiative QED effect.
 This would raise legitimate doubts about the wisdom of 
choosing $\mu_{\mbox{\tiny{anom}}}$ as the magnetic moment associated with the Born ring.
 To alleviate such worries, we report that we were able to reproduce our Bohr--Sommerfeld results with the factor $\alphaS^3$ 
in the magnetic interaction term in \refeq{eq:U} replaced by $\alphaS^{1+2k}$, and with the radius $R$ of Born's ring electron 
replaced by $0.49597832375\alphaS^{1+k}$, for $k\in\{0,1,2,3\}$ --- note the scaling law!
 Thus, as long as the electron has a tiny structure which supports a tiny permanent magnetic moment, there is a low-lying 
quantum-mechanical ground state of Ps, and the natural demand that it have near zero energy yields a relationship between 
the size of the structure and the magnitude of the magnetic moment it supports.\footnote{It should be clear that similar results
  will be obtained with the BLTP vacuum law. 
 Incidentally, since Bopp's $\varkappa$ plays a role analogous to a UV cutoff for QED \cite{FeynmanBOPP}, 
a QED type renormalization approach to remove the cutoff (i.e. sending $\varkappa\to\infty$) from our quantum-mechanical model in 
subsection \ref{sec:unPERTURBEDsqrtKG} will have to go hand in hand with sending $R\to 0$; but then, to prevent the Ps ground state energy
from going to $-\infty$ in this 
model, the magnetic coupling will have to be renormalized as well and sent to zero simultaneously.
 In the limit one will obtain point particles with no extra tiny magnetic moment (in addition to the Bohr magneton and the anomalous one), 
and in the limiting ground state configuration they would neutralize each other completely.
 This would be compatible with the annihilation scenario of QED, except that one may
let the ground state energy converge to a non-zero value in this limit.
 But since QED's UV cutoffs cannot be removed non-perturbatively, as it seems, one may take this as an indication that QED only
approximates the transition of Ps into its proper ground state by a mathematical annihilation operation, though very effectively so.\vspace{-15pt}}

 If confirmed by proper relativistic calculations, and by experiment, 
this novel ground state may mean that the universe is full of small mass, effectively bosonic particles which 
could form a large part of the mysterious dark matter in the universe. 
 It does not take much imagination to come up with the conjecture that the Ps spectrum suggested in this paper is representative
of the spectra of other atomic matter \&\ anti-matter systems, such as quarkonium etc. 
 Those systems likewise would have to be considered contributing to the dark matter in the universe.
 The upshot is: at the Big Bang (if there was a true singularity)
our universe may have started with a huge but finite number of particles in it, and all these may still be around 
(some of them hidden by the event horizon of a black hole). 
 We may already know what type of particles these are, all or most of them --- no need for speculative exotic matter models.
 And if dark matter is indeed mostly tightly bound particle \&\ anti-particle pairs, it would mean that the apparent total lopsidedness
of matter over anti-matter in the visible universe (see \cite{CERNb}) may be a misleading appearance. 
 The universe could be filled in almost equal amounts with matter and with anti-matter --- but only 
the tiny imbalance is precisely what is visible to us.
\bigskip

\noindent
\textbf{ACKNOWLEDGEMENT:} I sincerely thank Shadi Tahvildar-Zadeh and Shelly Goldstein 
for their encouragement and helpful comments. 
 I also thank J\"urg Fr\"ohlich for his encouragement at an earlier stage of this project (2004).
 After a first version of this paper was made public, I received many valuable comments;
especially from Detlev Buchholz, Eric Carlen, Elliott Lieb, Remo Ruffini, and Avy Soffer, for
which I am grateful.

%%%%%%%%%%%%%%%%%%%%%%%%%%%%%%%%%%%%%%%%%%%%%%%%%%%%%%%%%%%%%%
%%%%%%%%%%%%%%%%%%%%%%%%%%%%%%%%%%%%%%%%%%%%%%%%%%%%%%%%%%%%%%
%%%%%%%%%%%%%%%%%%%%%%%%%%%%%%%%%%%%%%%%%%%%%%%%%%%%%%%%%%%%%%
                \section*{Appendix}\label{sec:APP}\vspace{-5pt}
%%%%%%%%%%%%%%%%%%%%%%%%%%%%%%%%%%%%%%%%%%%%%%%%%%%%%%%%%%%%%%
%%%%%%%%%%%%%%%%%%%%%%%%%%%%%%%%%%%%%%%%%%%%%%%%%%%%%%%%%%%%%%
%%%%%%%%%%%%%%%%%%%%%%%%%%%%%%%%%%%%%%%%%%%%%%%%%%%%%%%%%%%%%%

 In this paper we have argued that electrons and anti-electrons may not annihilate each other, and we have suggested that similar
conclusions may be reached for other (ferminionic) particle \&\ anti-particle systems, such as quarkonium. 
 We have explained that what may appear like an annihilation may simply be a transition to a tightly bound state of such a 
particle \&\ anti-paarticle system.

 But what about the photon? 
 Photons surely get created and annihilated, don't they? 
 Not necessarily! 
 Of course, if one literally defines a photon, as is done in QED, as ``a spin-1 particle with zero rest mass and momentum $\hbar\kV$,'' 
then as soon as the momentum of the photon changes from $\hbar \kV$ to $\hbar\kV'$, ``the original $\kV$ photon'' is gone, and 
``a $\kV'$ photon'' has appeared. 
 But this is not distinguishable from the scenario in which a photon is really a particle, perhaps a point particle, which simply
changes its momentum from $\hbar \kV$ to $\hbar\kV'$, very much as envisaged by Einstein, see \cite{KTZphoton}.
 It is quite conceivable, and logically perfectly admissible, that the emission / absorption of photons by atoms does not mean that
photons are created / annihilated in this process, but simply ejected / captured.
 In a forthcoming paper \cite{KLTZ} it will be demonstrated that this scenario is indeed happening in 
a relativisitic photon-electron model in 1+1 spacetime dimensions.
 Of course, a 1+1 dimensional model calculation is only a first baby step, but also quantum field theory once started in
this way.  
 At the very least, it suggests to be open-minded about the possibility that annihilation / creation of photons only appears
to be happening in nature, and that it may be as misleading as the apparent  annihilation / creation of Ps. 
\vspace{-10pt}
%%%%%%%%%%%%%%%%%%%%%%%%%%%%%%%%%%%%%%%%%%%%%%%%%%%%%%%%%%%%%%%%%%
%%%%%%%%%%%
%%%%%%%%%%% And finally: The bibliography 
%%%%%%%%%%%
%%%%%%%%%%%%%%%%%%%%%%%%%%%%%%%%%%%%%%%%%%%%%%%%%%%%%%%%%%%%%%%%%%

\end{document}